# Single mode output by controlling the spatiotemporal nonlinearities in mode-locked femtosecond multimode fiber lasers

Uğur Teğin[1,2,*], Babak Rahmani[2], Eirini Kakkava[1], Demetri Psaltis[1], and Christophe Moser[2]

[1]Optics Laboratory, École Polytechnique Fédérale de Lausanne, Switzerland
[2]Laboratory of Applied Photonics Devices, École Polytechnique Fédérale de Lausanne, Switzerland
*e-mail: ugur.tegin@epfl.ch

## Abstract

The performance of fiber mode-locked lasers is limited due to the high nonlinearity induced by the spatial confinement of the single-mode fiber core. To massively increase the pulse energy of the femtosecond pulses, amplification is obtained outside the oscillator. Recently, spatiotemporal mode-locking has been proposed as a new path to fiber lasers. However, the beam quality was either highly multimode and/or with low pulse energy. Here we present an approach to reach high energy per pulse directly in the mode-locked multimode fiber oscillator with a near single-mode output beam. Our approach relies on spatial self-beam cleaning via the nonlinear Kerr effect and we demonstrate a multimode fiber oscillator with $M^2$<1.13 beam profile, up to 24 nJ energy and sub-100 fs compressed duration. The reported approach is further power scalable with larger core sized fibers and could benefit applications that require high power ultrashort lasers with commercially available optical fibers.

## Introduction

Fiber laser dynamics have been studied extensively in the past decades to generate femtosecond pulses with high energies and peak powers [1]. Numerous laser designs are developed to understand nonlinear wave propagation under partial feedback conditions. By tuning complex cavity dynamics in single-mode fiber cavities, self-organization of longitudinal cavity modes with various temporal profiles and central wavelengths have been realized such as soliton [2], similariton [3,4] and dissipative soliton [5,6]. Due to the high spatial confinement in the small single mode fiber core, nonlinear effects appear at moderate peak power in solid cores and the accumulation of excessive nonlinear phase leads to pulse breakup, which then limits the achievable pulse energies .

To overcome this limitation, custom-made sophisticated fibers, having large single mode areas in a photonic crystal fiber (PCF), were proposed to reach the µJ pulse energy level [7,8]. The fiber used in these demonstrations with its mode field diameter of 70 um needed to be kept straight to avoid bending losses and ensure stability. They share the same limitations as solid state lasers in the sense that they are rigid and cannot be spliced with conventional techniques. As a result such lasers do not share the features of fiber lasers that render them advantageous in practice.

Recently, spatiotemporal mode-locking has been demonstrated in commercially available MMF cavities with graded-index multimode fibers (GRIN MMFs) by harnessing their low modal dispersion and inherent periodic self-focusing to produce a coherent superposition of transverse and longitudinal modes in an all-normal dispersion regime [9,10,11]. These studies presented cavities with dissipative soliton pulse operation with low output beam quality. In a recent study, improvement in the output beam profile (output beam $M^2$ value <1.4 ) of a spatiotemporal mode-locked fiber laser was reported in an amplifier similariton pulse leading to pulse energies of 2.4 nJ [12].

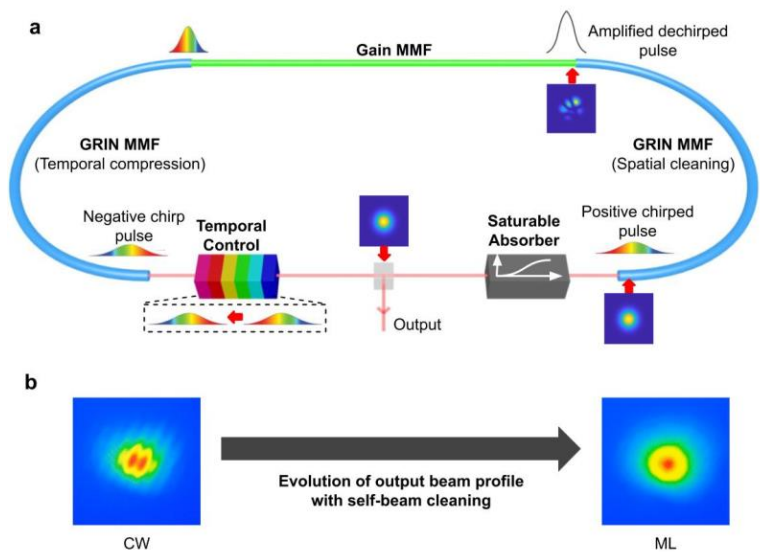

**Figure 1 Conceptual outline of the multimode fiber cavity and schematic of spatiotemporal mode-locking with self-beam cleaning. a** Conceptual outline of dispersion-managed cavity design with indicated temporal dynamics. Laser pumping scheme is not presented. **b** Schematic of mode-locking mechanism and experimentally measured output beam profile evolution. CW, continuous wave; ML, mode-locked.

Here we demonstrate an approach to generate >20 nJ, sub-100 fs pulses with near Gaussian output beam shape by controlling the spatiotemporal cavity dynamics of multimode fiber lasers. This pulse energy represents a tenfold improvement over previously reported MMF oscillators with Gaussian beam shape. Moreover, our method is limited only by the damage threshold of the fiber, splices and can be increased to the µJ energy level. The essence of the approach relies on the experimental observation that a high-intensity pulse with a multimode beam profile can be transformed into a near-Gaussian beam by Kerr-induced self-beam cleaning in a GRIN MMF [13]. The optical wave propagation inside the cavity is designed to achieve nonlinear beam cleaning with spatiotemporal mode-locking. With our approach, spatiotemporally mode-locked fiber lasers overcome the power limitations of mode-locked single-mode fiber lasers without sacrificing the output beam quality. Moreover, our approach is not limited to the demonstrated power levels and is scalable with standard large core multimode fibers.

We experimentally demonstrate that the highly multimode beam profile observed at the output of a continuous wave multimode fiber cavity is transformed to a stable Gaussian beam profile when the oscillator is spatiotemporally mode-locked. Our numerical studies verified that there is an energy exchange from higher-order modes to lower order modes in the propagating GRIN MMF section of the laser cavity for experimentally reached high power level. Inside the spatiotemporal mode-locked cavity, Kerr-induced self-beam cleaning creates a minimum loss condition to the emerging mode-locked pulses. The reported multimode fiber laser generates sub-100 fs pulses with high pulse energy (>20 nJ) and good beam quality of $M^2$ value is less than 1.13.

# Results

## Numerical Studies

Our approach is inspired by dispersion-managed mode-locking in single mode lasers which is a method that was initially proposed to overcome the power limitation of soliton mode-locked lasers [14, 15]. Inside a dispersion-managed cavity, the pulse stretches and compresses significantly every round trip while having a near zero net-cavity group-velocity dispersion (GVD). This approach leads to the generation of ultrashort pulses (sub-100 fs) with broad-spectrum and high energy per pulse [16,17].

In our study, the dispersion-managing approach is adapted to a multimode fiber cavity with a similar motivation, to generate intracavity short pulses with high peak powers sufficient to trigger strong spatiotemporal interactions to effect spatial beam clean-up The proposed cavity is illustrated in Fig. 1a. Silica fibers feature positive GVD for the emission wavelengths of the Yb-fiber and stretch the propagating pulse with positive chirp. In our design, a grating pair is placed as a dispersion-balancing section to provide negative GVD values and change the sign of the chirp on the pulse. Such a temporal change causes compression of the propagating pulses in the following fiber sections (GRIN MMF and gain MMF).

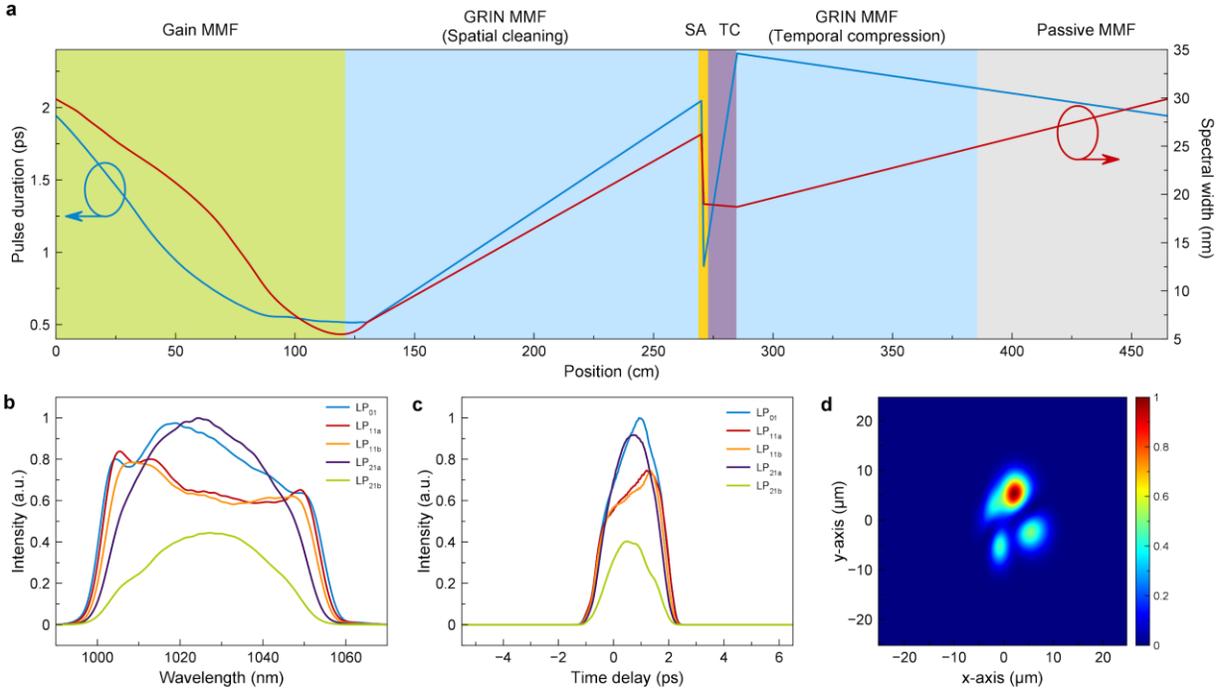

**Figure 2 Pulse evolution in the laser simulations. a** Simulated pulse duration and spectral bandwidth variation over the cavity: SA, saturable absorber; PC, pulse compressor. **b** Simulated mode-resolved spectral intensity. **c** Simulated mode-resolved pulse intensity. **d** Numerically obtained mode-locked beam profile.

We used the multimode nonlinear Schrödinger equation to numerically simulate the multimode laser to analyze pulse propagation dynamics and determine cavity parameters. A stable mode-lock regime after a few round-trips around net cavity GVD $\beta^{(2)}_{net}$ = 0.013 ps$^2$ is achieved and dispersion-managed soliton pulse formation is numerically obtained with various powers and cavity excitation conditions. An example of the evolution of the intracavity pulse in one round trip is illustrated in Fig. 2a as a function of position inside the multimode cavity. By engineering the cavity dynamics, the shortest pulse duration is achieved at the end of the gain fiber section where the amplification is maximum such that the highest intracavity peak power is obtained in the GRIN MMF section. This unique design can allow pulses to reach the Kerr-induced beam cleaning threshold inside the cavity. In our simulations, the pulse experiences more than 6 times spectral broadening in one round-trip. The temporal profile and spectrum of the output pulse measured at the nonlinear polarization evolution (NPE) port, are presented in Fig. 2b and Fig. 2c. For the simulated cavity, mode-locked pulses with 5.1 nJ energy and 2.12 ps duration Gaussian-like temporal shape are generated from quantum noise. Numerically obtained output beam profile with the considered modes is presented in Fig. 2d. For further details and simulations with different multimode fiber excitation conditions, see Methods and Supplementary Information I.

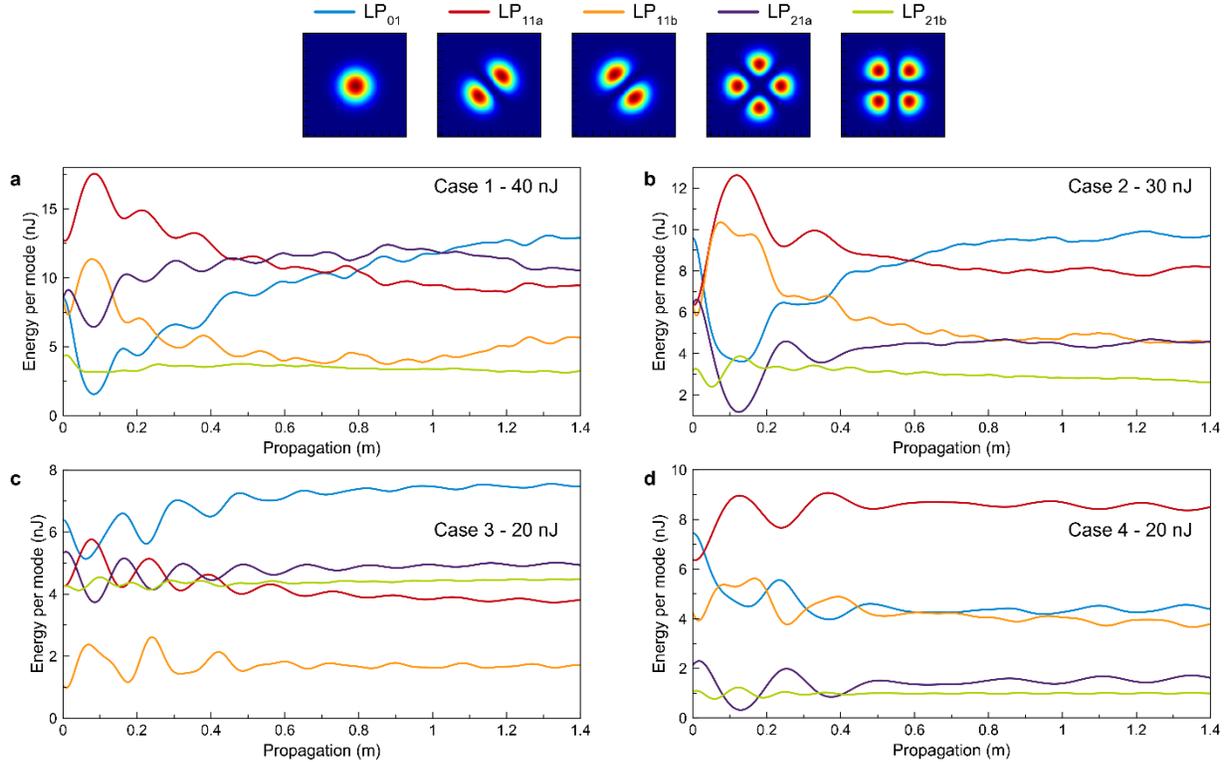

**Figure 3 Numerical investigation of Kerr-induced self-beam cleaning inside GRIN MMF.** Propagation of 450 fs pulse with **a** 40 nJ pulse energy and [20%, 30%, 20%, 20%, 10%] initial coupling condition, **b** 30 nJ pulse energy and [30%, 20%, 20%, 20%, 10%] initial coupling condition, **c** 20 nJ pulse energy and [30%, 20%, 5%, 25%, 20%] initial coupling condition and **d** 20 nJ pulse energy and [35%, 30%, 20%, 10%, 5%] initial coupling condition.

To investigate the possibility of the Kerr-induced self-beam cleaning in the cavity, single-pass numerical simulations were performed with high numerical accuracy. The energy exchange behavior between the modes is presented in Fig. 3. The mode-locking simulations demonstrated in Fig. 2 suggest ~450 fs pulse duration after the gain section and we numerically investigate the effect of intracavity pulse energy to energy exchange between the modes of GRIN MMF with different initial excitation scenarios. For the excitation case applied in mode-locking simulations between the simulated modes, when the pulse energy reaches to 20 nJ, energy fluctuations between the simulated modes starts to decrease. For a 30 nJ pulse the fundamental mode starts to increase its energy content and for 40 nJ pulse it becomes dominant at the end of the GRIN MMF section of the cavity (see Fig. 3a). For different excitation cases, the required pulse energy to achieve similar modal interaction is observed to be lower as it is shown in Fig. 3b and Fig. 3c. For some cases, we observed that instead of the fundamental mode, an alternative low order mode such as LP11b is observed as a leading mode at the end of GRIN MMF for 20 nJ pulse energy (see Fig. 3d). Contrary to single-pass propagation, inside a laser cavity, a small improvement in beam shape can accumulate and a steady-state beam cleaning can be achieved after multiple roundtrips. The aforementioned numerical studies indicate that the pulse energy required for intracavity Kerr-induced beam cleaning is within reach of the designed dispersion-managed multimode cavity.

## Experimental Studies

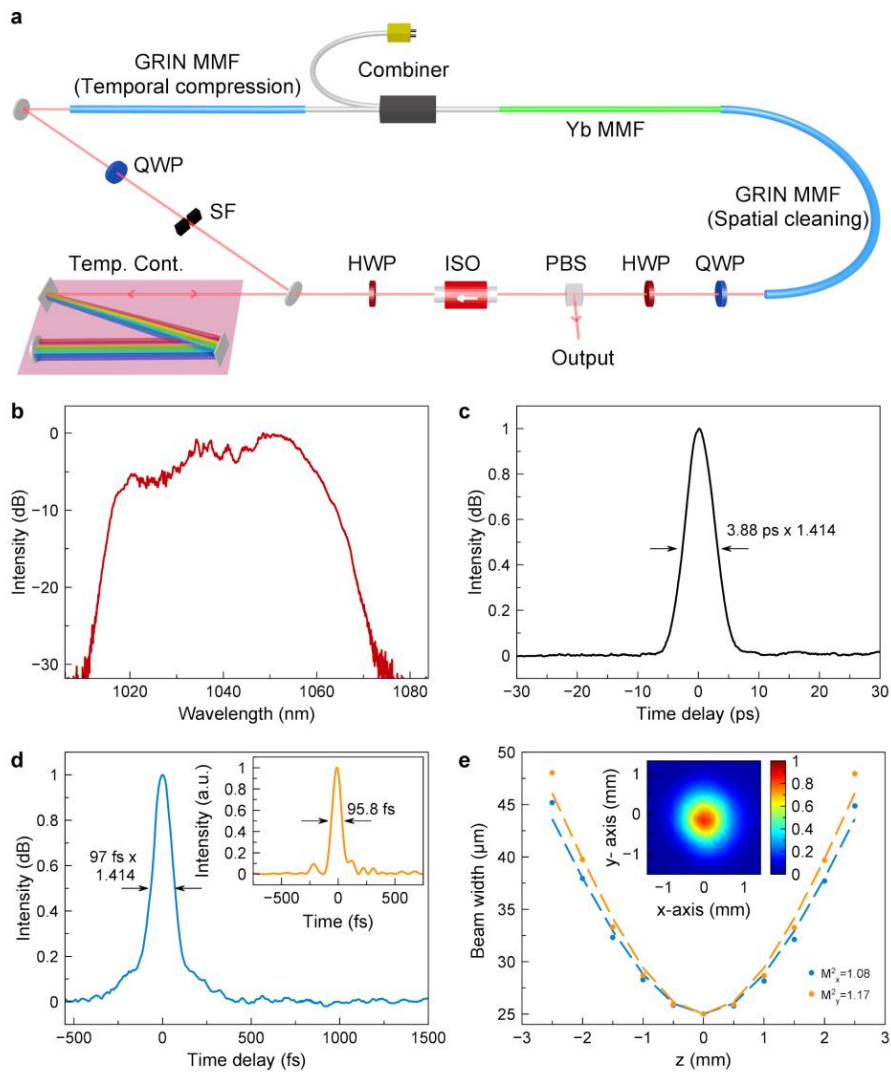

**Figure 4 Schematic of the laser cavity and experimental measurements. a** QWP, quarter-wave plate; HWP, half-wave plate; PBS, polarizing beam splitter; ISO, isolator; SF, spatial filter. **b** Measured mode-locked spectrum. **c** Intensity autocorrelation trace of the chirped pulse. **d** Intensity autocorrelation trace of the dechirped pulse and PICASO-retrieved dechirped pulse shape (inset). **e** Measured beam profile for mode-locked operation and M2 measurement of the beam profile.

Guided by the simulation results, the experimental cavity, presented in Fig.4, is constructed and studied. The intracavity grating compressor is tuned to provide -0.0987 ps$^2$ GVD to compensate 0.1116 ps$^2$ GVD of the fiber sections of the cavity each round trip. Single-pulse mode-locking is easily achieved by adjusting cavity polarization by the wave plates with a repetition rate of 36 MHz. A major improvement from a highly multimode beam profile to a Gaussian-like symmetric beam profile is observed at the laser output when the operation regime changed from continuous-wave to mode-locked [see Fig. 1b and Supplementary Video 1]. The measured Gaussian-like beam profile remains similar when pulse energy is increased up to 24 nJ by gradually increasing the pump power level. Above 24 nJ output pulse energy, secondary pulse formation is observed inside the cavity as a power limiting factor for the single-pulse operation regime.

We performed a detailed characterization of the laser when the pump power is set to 3.5 W where the laser generates 20 nJ pulses. As expected for a dispersion-managed cavity, pulse spectra with large bandwidth (~40 nm) are measured for this power level [see Fig. 4 b]. The pulse duration of the chirped pulses is measured with second-order intensity autocorrelation as 3.88 ps with the Gaussian deconvolution factor of 1.414 as shown in Fig. 4 c. These chirped pulses are later dechirped (compressed) by an external grating compressor with a diffraction grating pair to 97 fs (see Fig. 4 d). To determine the temporal profile of the pulses from the measured spectrum and autocorrelation data, the PICASO algorithm is employed [18]. The resulting pulse shape features 95.8 fs pulse duration with a Gaussian profile and presented in Fig. 4 d – Inset. No secondary pulse formation or periodic oscillation of the pulse train is observed (see Supplementary Information II). As presented in Fig. 4 e, the beam has a symmetric shape and to determine its quality, M2 measurements are performed. For 20 nJ output pulse energy, average $M^2$ is measured as 1.13 ($M^2_x$ = 1.08 and $M^2_y$ = 1.17).

## Discussion

To investigate the effect of intracavity pulse energy to output beam profile in detail, we performed $M^2$ measurements to pulses with different energy by adjusting pump power. The changes in the output beam quality are challenging to differentiate from the near-field beam profiles but M2 measurements provide more significant information. In our measurements, we observed that $M^2$ value is increasing for low pulse energy and we measured $M^2$ = 1.85 for 4 nJ and $M^2$ = 1.48 for 7.2 nJ (see Supplementary Information II). When the $M^2$ value for the presented high pulse energy is also considered, this measurement is the experimental proof of the intracavity Kerr-induced beam cleaning in our design.

To understand the beam cleaning mechanism, the effect of the cavity alignment condition on the mode-locked beam profile is investigated. Differences in the free-space orientation of the laser are achieved by changing the cavity alignment and the position of the intracavity spatial filter. We observed that the presented beam-profile remains similar for different alignment configuration although the mode-locked spectrum and pulse energy changes (see Supplementary Information II for examples). A comparison of the predicted beam profile by the simulation (Fig. 2d) reveals that experimentally we obtain a much better beam shape (Fig. 4e). One of the reasons for this is the limited number of modes in the simulation and the low pulse energy that was simulated. In the single-pass simulations reported in Fig. 3, we found that with the higher pulse energy beam cleaning improves which is consistent with the experimental result. Also, given the fact that the laser pulse oscillates inside the cavity, a slight improvement in the beam profile due to the Kerr effect during each round trip, enhanced self-beam cleaning is expected.

The differences between the supported number of modes by the multimode fibers used in the spatiotemporally mode-locked laser causes mode dependent loss to the propagating field. For spatiotemporal mode-locking with the presented engineered cavity approach, Kerr-induced beam cleaning can cause a minimum loss condition inside the multimode fiber cavity, similar to the Kerr-lens realization of solid-state lasers. Based on our results, one can explain the measured output beam profile evolution in the laser output with the minimum loss principle [19]. The effect of Kerr-induced beam cleaning to cavity gain/loss dynamics needs to be further investigated in future work.

In conclusion, we reported a multimode fiber laser design with intracavity Kerr-induced self-beam cleaning to realize high energy, ultrashort pulses with good beam quality. By engineering nonlinear intracavity propagation of the mode-locked pulses, we numerically and experimentally demonstrated a multimode cavity design with Kerr-induced self-beam cleaning. The presented cavity dynamics show that engineered intracavity temporal pulse properties enable a route to generate high beam quality when mode-locking is achieved. For various alignment orientations and pulse energy, drastic improvement of the output beam profile is experimentally reported when mode-locking is achieved. The presented oscillator generates sub-100 fs pulses with >20 nJ pulse energy while exhibiting good beam quality of $M^2$ value is less than 1.13. The combination of good beam quality, high pulse energy and sub-100 fs pulse duration from a fiber laser consists of commercially available, standardized components is a promising platform for various laser-related fields. The presented technique can be easily adapted to fibers with a larger core size to increase pulse energy while preserving single-pulse operation with sub-100 fs durations.

## Methods

**Multimode oscillator simulations.** Numerical simulations for mode-locked pulse formation are conducted for the model used by Tegin et al. [12]. For GRIN MMF sections of the cavity, a multimode non-linear Schrödinger equation [20] is solved by considering the five linearly polarized (LP) modes (more details of simulations can be found in Supplementary Information I). Simplifications such as simulating few-mode fiber sections as single-mode to decrease the computation time and defining coupling ratios before and after the GRIN MMF sections to mimic the effect of splice points are performed. Simulations are initiated with quantum noise fields. The integration step for GRIN MMF sections is defined as the ratio of the self-imaging period of the fiber with 4. For the simulation result shown in Fig. 1, the gain fiber is modeled with Lorentzian gain shape with 50 nm bandwidth, 30 dB small-signal gain and 3.2 nJ saturation energy. The coupling condition between the few-mode gain fiber and GRIN MMF is simulated as [20%, 30%, 20%, 20%, 10%]. The intracavity spatial filtering is applied to considered modes by allowing their transmission with [50%, 50%, 50%, 50%,0%]. The coupling for the field propagating from the GRIN MMF to a few-mode fiber is calculated as a summation of modes with the transmission coefficients [100%, 100%, 100%, 0%, 0%].

**Single-pass beam cleaning simulations.** The simulations regarding beam cleaning are performed by numerically solving a multimode non-linear Schrödinger equation with high numerical accuracy and small integration step. To achieve this goal, the integration step is defined as the ratio of the self-imaging period of GRIN MMF with 20 and the fourth-order Runge-Kutta in the Interaction Picture method is used for accuracy [21]. (Further details can be found in Supplementary Information I)

**Experiments.** A 20 W, 976 nm pump diode (II-VI Photonics BMU20-976S-01-R) is spliced to the oscillator cavity with a pump combiner (Lightel MPC (2+1) x1) compatible with a 1.3 m MMF gain section (nLight Yb-1200 10/125). To excite the higher-order modes of the GRIN MMF (Thorlabs GIF50C), the gain fiber is spliced to the 1.4 m GRIN MMF with a small offset (5 µm). After the first GRIN MMF section, light is collimated and travels through wave plates, polarizing beam splitter, isolator, grating pair (600line/mm) and spatial filter (randomly placed pinhole). With wave plates and a polarizing beam splitter, NPE is implemented as an artificial saturable absorber. The additional half-wave plate is placed between the isolator and the grating pair to control the first-order reflection efficiency of the gratings. The free-space ends of the GRIN MMFs are angle-cleaved to eliminate parasitic back reflections. Another GRIN MMF section with 1 m length establishes the ring cavity loop. Self-starting mode-locking was achieved by adjusting the intracavity wave plates. Spectrum measurements are performed with an optical spectrum analyzer (Ando AQ6317B) and spectrometer (Ocean Optics HR4000-CG-UV-NIR). Beam profile measurements are performed with CMOS cameras (Edmund Optics EO-32121M and Thorlabs DCC1545M). Temporal pulse measurements are performed with an autocorrelator (Femtochrome FR-103).

*Supplementary information*

# Single mode output by controlling the spatiotemporal nonlinearities in mode-locked femtosecond multimode fiber lasers

Uğur Teğin[1,2], Babak Rahmani[2], Eirini Kakkava[1], Demetri Psaltis[1], and Christophe Moser[2]

[1]Optics Laboratory, École Polytechnique Fédérale de Lausanne, Switzerland
[2]Laboratory of Applied Photonics Devices, École Polytechnique Fédérale de Lausanne, Switzerland
*e-mail: ugur.tegin@epfl.ch

**Supplementary Discussion I:**

**Multimode oscillator simulations**

Numerical simulations for mode-locked pulse formation are conducted for the model used by Tegin et al. [12]. Few mode fiber sections are considered as single-mode and coupling between these and GRIN MMF sections are approximated with coupling coefficients used under conservation of energy principle. The number of modes in GRIN MMF fibers is simplified as the first 5 linearly polarized (LP) modes ($LP_{01}$, $LP_{11a}$, $LP_{11b}$, $LP_{21a}$, $LP_{21b}$). All spatiotemporal mode-locking simulations are initiated with quantum noise fields. The gain segment is modeled with Lorentzian shape and 30 dB small-signal gain. Nonlinear polarization evolution (NPE) saturable absorber is modeled as presented in Eq.1.

$$T = q_0 + q_1 sin^2\left(\frac{I}{I_{sat}}\right) \qquad (1)$$

Here, $q_0$ is the minimum transmission, $q_1$ is the modulation depth, $I$ is the instantaneous pulse power and $I_{sat}$ is the saturation power. To decrease computation time, the integration step is defined as equal to the self-imaging period of GRIN MMF and Raman scattering term is not considered in the spatiotemporal mode-locking simulations.

For the mode-locked multimode simulations shown in Fig. S1 and Fig. S2, laser cavities experience net cavity GVD $\beta^{(2)}_{net}$ = 0.013 ps² and spatial filtering coefficients are similar to the oscillator presented in Fig. 2. Spatiotemporal mode-locking achieved with 3.1 nJ gain saturation energy, free space coupling loss 0.6 and [30%, 20%, 20%, 20%, 10%] initial coupling condition between gain MMF and GRIN MMF for the results presented in Fig. S1. The mode-locked pulses with 4.1 nJ energy and 1.96 ps duration Gaussian-like temporal shape is generated. By changing the intracavity coupling condition to [35%, 30%, 20%, 10%, 5%], free space coupling loss to 0.5 and adjusting the small signal gain to 2.25 nJ, spatiotemporal mode-locking with 2.4 nJ energy and 1.72 ps duration is achieved as shown in Fig. S2.

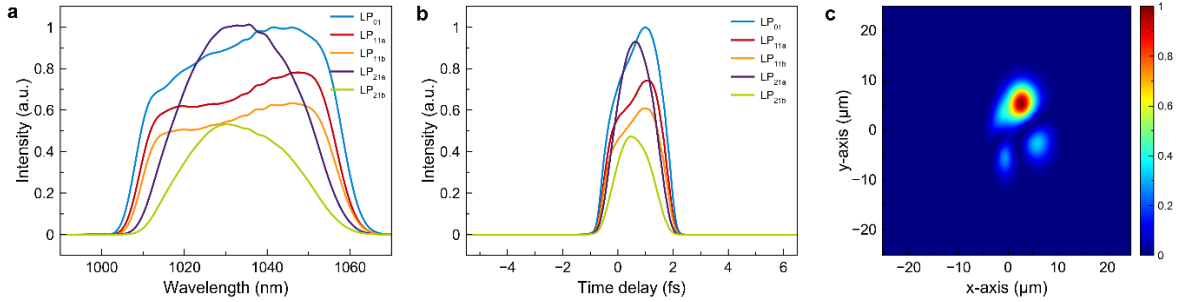

**Figure S1 Simulated mode-locked multimode laser with [30%, 20%, 20%, 20%, 10%] initial coupling condition between gain MMF and GRIN MMF. a** Simulated mode-resolved spectral intensity. **b** Simulated mode-resolved pulse intensity. **c** Numerically obtained mode-locked beam profile.

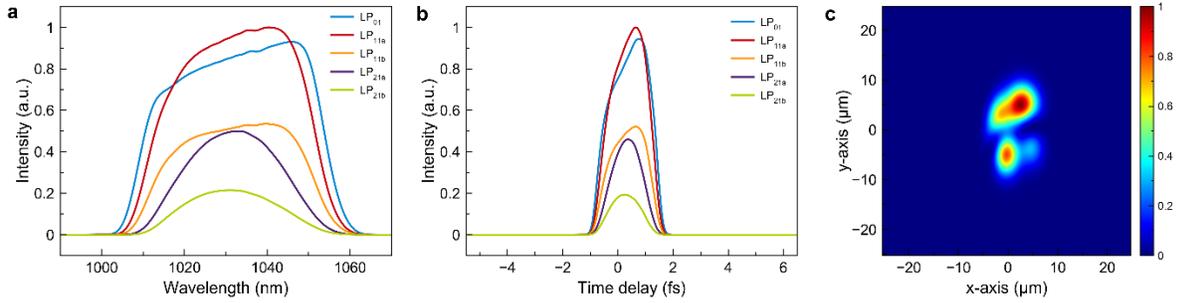

**Figure S2 Simulated mode-locked multimode laser with [35%, 30%, 20%, 10%, 5%] initial coupling condition between gain MMF and GRIN MMF. a** Simulated mode-resolved spectral intensity. **b** Simulated mode-resolved pulse intensity. **c** Numerically obtained mode-locked beam profile.

**Single-pass beam cleaning simulations**

The multimode nonlinear Schrödinger equation with a Raman scattering term (Eq. 2) is numerically solved with high numerical accuracy and small integration steps to investigate modal interactions inside the GRIN MMF segment after the gain fiber.

$$\frac{\partial A_p}{\partial z}(z,t) = i\delta\beta_0^{(p)}A_p - i\delta\beta_1^{(p)}\frac{\partial A_p}{\partial t} - i\frac{\beta_2}{2}\frac{\partial^2 A_p}{\partial t^2} + \frac{\beta_3}{6}\frac{\partial^3 A_p}{\partial t^3}$$
$$+ i\gamma \sum_{l,m,n} \eta_{plmn}\left[(1-f_R)A_l A_m A_n^* + f_R A_l \int h_R A_m(z,t-\tau)A_n^*(z,t-\tau)d\tau\right] \quad (2)$$

Here $\eta_{plmn}$ is the nonlinear coupling coefficient, $\delta\beta_0^{(p)}$ ($\delta\beta_1^{(p)}$) is the difference between first (second) Taylor expansion coefficient of the propagation constant for corresponding and the fundamental mode and $f_R$ is the fractional contribution of the Raman effect, $h_R$ is the delayed Raman response function. Relative index difference is assumed to be 0.01. For numerical integration, a used a fourth-order Runge-Kutta in the Interaction Picture method [21] is employed and integration steps are defined as the ratio of the self-imaging period of the simulated GRIN MMF with 20. Pump pulses centered at 1030 nm with 450 fs duration are numerically propagated for 1.4 m distance. For each scenario with different initial coupling conditions, the effect of pulse energy to modal energy exchange is studied (see Fig. S3, Fig. S4, Fig. S5, Fig. S6). In our simulations, permanent energy transfer to fundamental mode (LP01), Kerr-induced self-beam cleaning, is reported for the considered cases. The changes in the initial coupling condition affect the threshold power to initiate Kerr-induced self-beam cleaning.

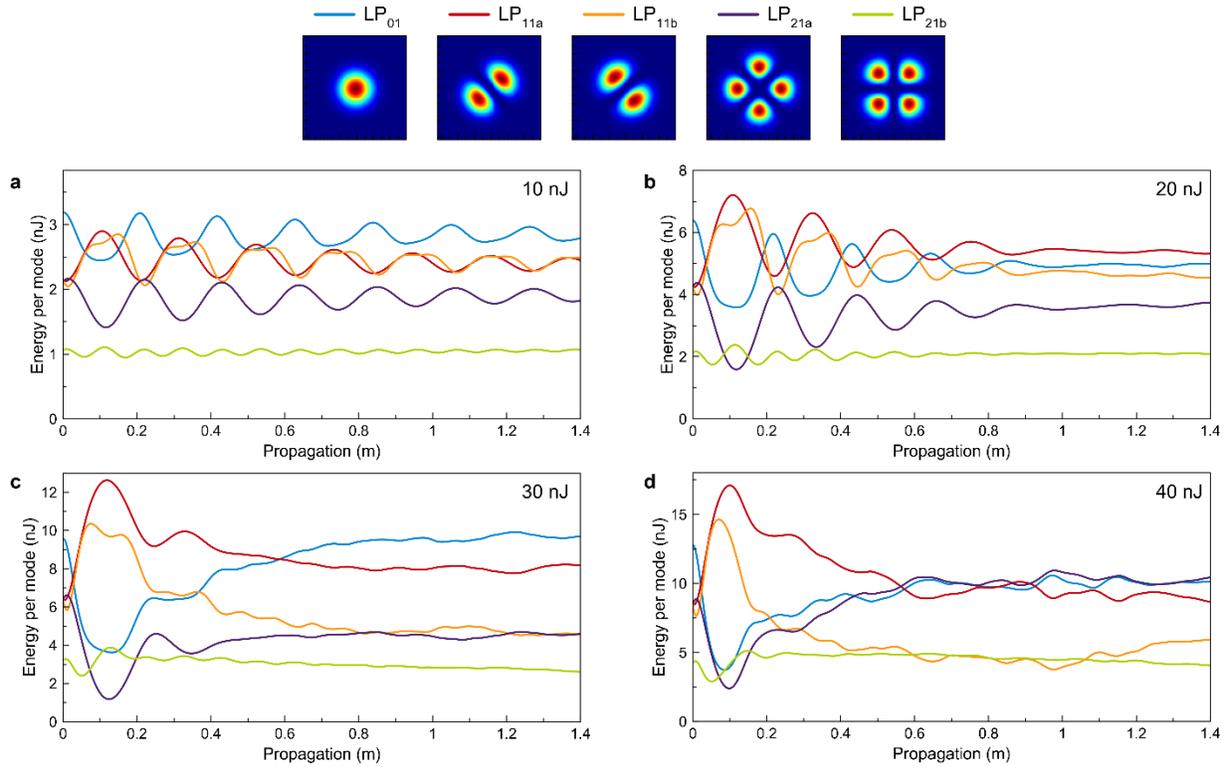

**Figure S3 Numerical investigation of Kerr-induced self-beam cleaning inside GRIN MMF for [30%, 20%, 20%, 20%, 10%] initial coupling condition. a** 10 nJ pulse energy, **b** 20 nJ pulse energy, **c** 30 nJ pulse energy and **d** 40 nJ pulse energy.

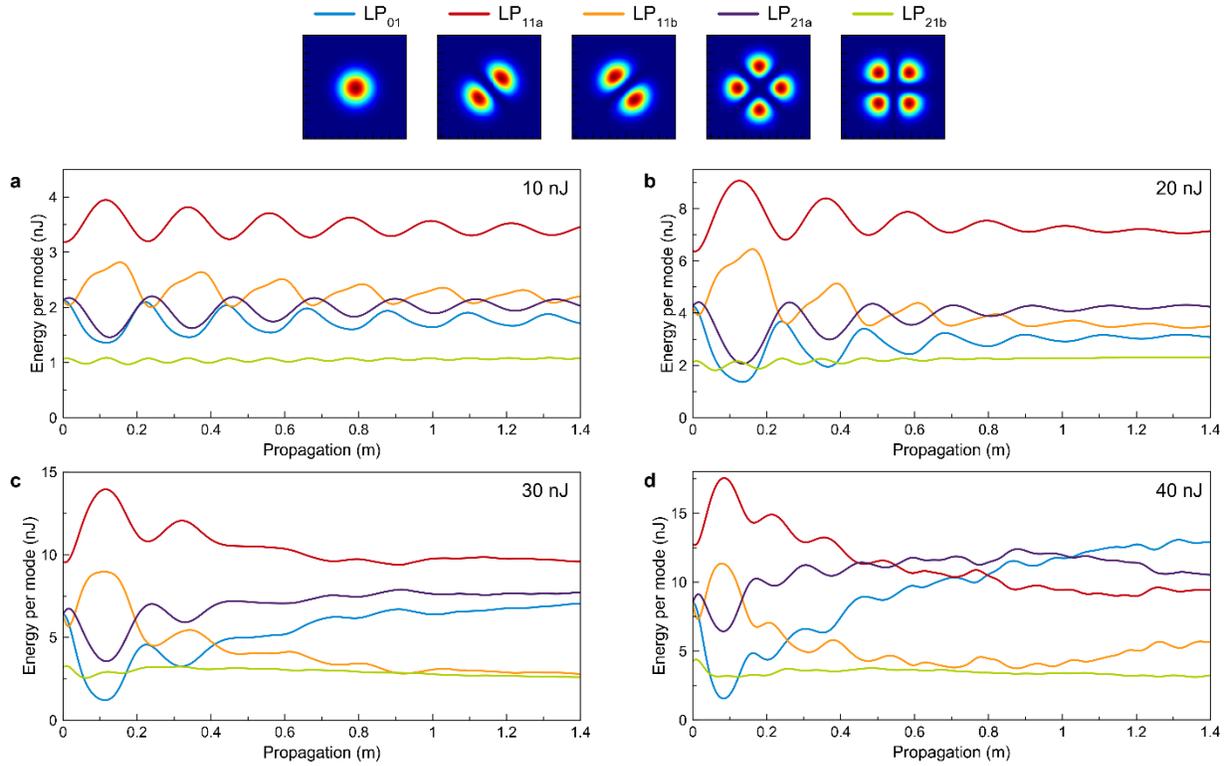

**Figure S4 Numerical investigation of Kerr-induced self-beam cleaning inside GRIN MMF for [20%, 30%, 20%, 20%, 10%] initial coupling condition. a** 10 nJ pulse energy, **b** 20 nJ pulse energy, **c** 30 nJ pulse energy and **d** 40 nJ pulse energy.

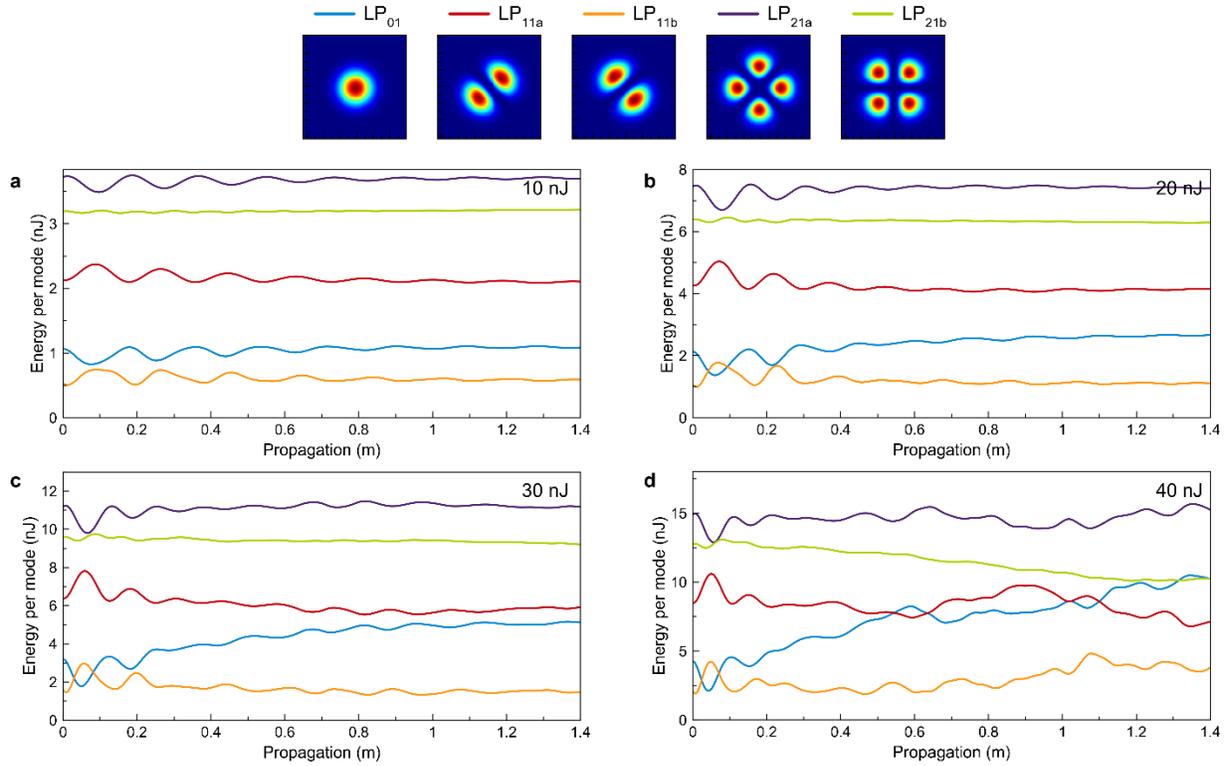

**Figure S5 Numerical investigation of Kerr-induced self-beam cleaning inside GRIN MMF for [10%, 20%, 5%, 35%, 30%] initial coupling condition. a** 10 nJ pulse energy, **b** 20 nJ pulse energy, **c** 30 nJ pulse energy and **d** 40 nJ pulse energy.

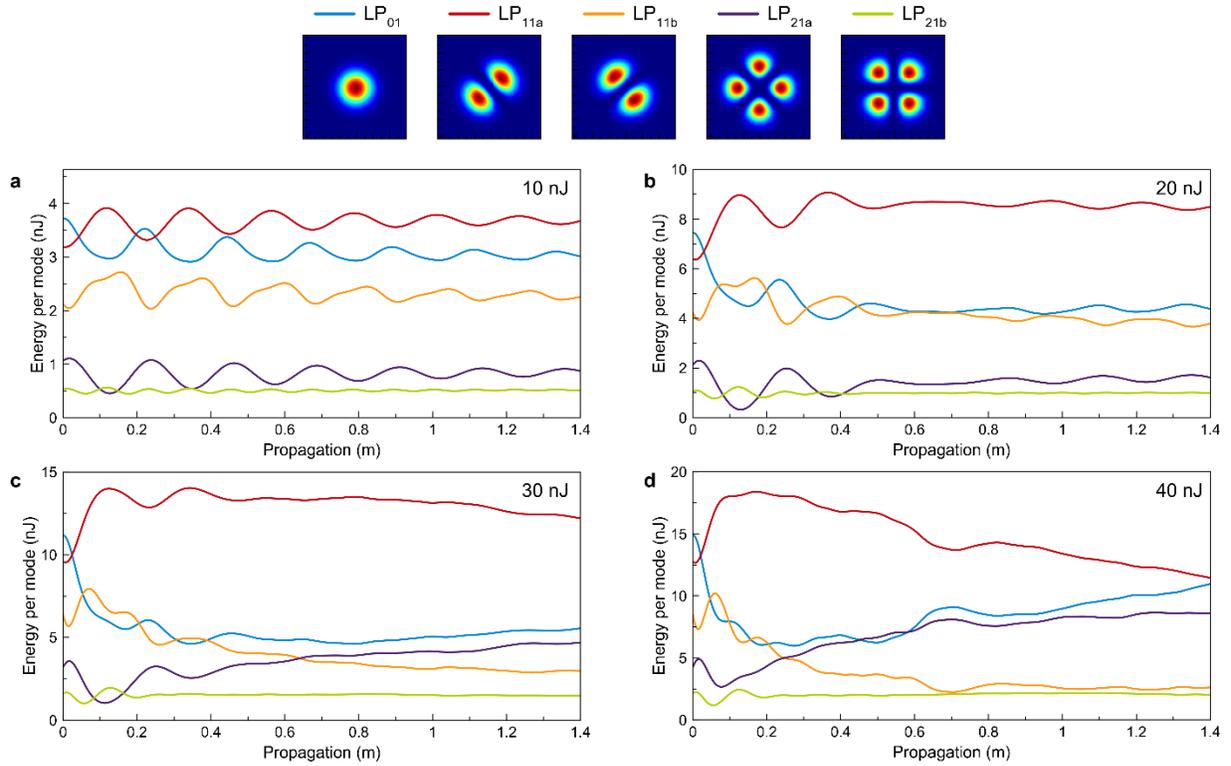

**Figure S6 Numerical investigation of Kerr-induced self-beam cleaning inside GRIN MMF for [35%, 30%, 20%, 10%, 5%] initial coupling condition. a** 10 nJ pulse energy, **b** 20 nJ pulse energy, **c** 30 nJ pulse energy and **d** 40 nJ pulse energy.

**Supplementary Discussion II:**

For a different alignment configuration of the cavity, spectral and the output beam profile changes are presented in Fig. S7, for incising pulse energy. A significant difference in the near-field output beam profile is not observed although spectral changes reported. In our measurements, pulse output pulse durations are observed as less chirped (<3.5 ps) and more compressible (<90 fs) for low pulse energies. Inside the multimode cavity, due to the aforementioned temporal changes, the peak power of the pulses after the gain fiber segment can be sustained above the threshold for intracavity self-beam cleaning.

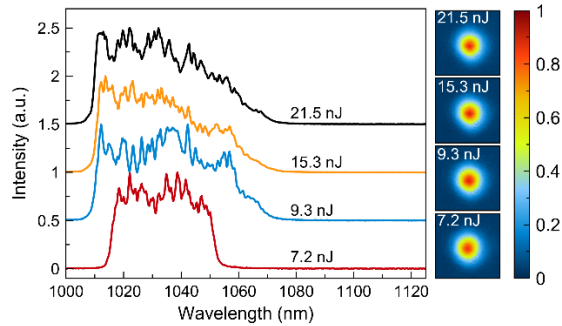

**Figure S7 Measurements for different cavity alignment configuration.** Spectral and spatial changes are presented for different pulse energy values.

Although significant changes in the output beam profile are not observed for different power levels, $M^2$ measurements are conducted to investigate the spatial changes in detail. For 4 nJ output pulse energy, an average $M^2$ value of 1.85 (Fig. S8 a) is recorded. When the output pulse energy is increased to 7.2 nJ, the average $M^2$ value is measured as 1.48 (see in Fig. S8 b).

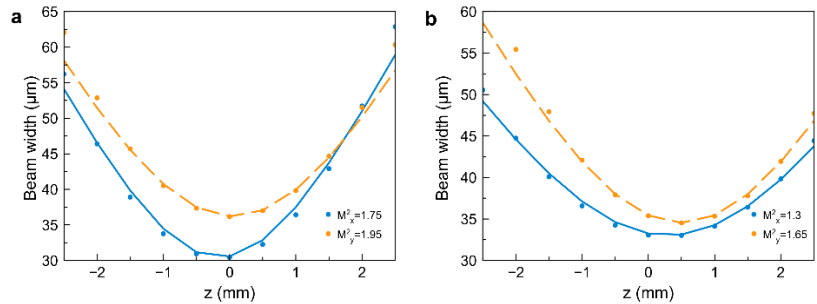

**Figure S8 $M^2$ measurements for different pulse energies. a** for 4 nJ pulse energy, **b** for 7.2 nJ pulse energy.